\documentclass[twocolumn]{aastex63}
\usepackage{bm}
\usepackage{amsmath,amsthm,amsfonts,amssymb,amscd}
\usepackage{lineno}
\received{xxxx xx, xxxx}
\revised{xxxx xx, xxxx}
\accepted{xxxx xx, xxxx}
\submitjournal{ApJ}

\shorttitle{P$^{3}$M lensing}
\shortauthors{Xu \& Jing}

\begin{document}

\title{An Accurate P$^{3}$M Algorithm for Gravitational Lensing Studies in Simulations}

\correspondingauthor{Yipeng Jing}
\email{ypjing@sjtu.edu.cn}

\author[0000-0002-7697-3306]{Kun Xu}
\affil{Department of Astronomy, School of Physics and Astronomy, Shanghai Jiao Tong University, Shanghai, 200240, China}

\author[0000-0002-4534-3125]{Yipeng Jing}
\affil{Department of Astronomy, School of Physics and Astronomy, Shanghai Jiao Tong University, Shanghai, 200240, China}
\affil{Tsung-Dao Lee Institute, and Shanghai Key Laboratory for Particle Physics and Cosmology, Shanghai Jiao Tong University, Shanghai, 200240, China}


\begin{abstract}
We present a two-dimensional (2D) Particle-Particle-Particle-Mesh (P$^3$M) algorithm with an optimized Green function and adaptive softening length for gravitational lensing studies in N-Body simulations. The analytical form of the optimized Green function $\hat{G}(\bm{k})$ is given. The softening schemes ($S$) are studied for both the PM and the PP calculations in order for accurate force calculation and suppression of the particle discreteness effect. Our method is two orders of magnitude more accurate than the simple PM algorithm with the {\it poor man's} Green function ($\propto1/k^2$) at a scale of a few mesh cells or smaller. The force anisotropy is also much smaller than the conventional PM calculation. We can achieve a force accuracy better than 0.1 percent at all scales with our algorithm, which makes it an ideal (accurate and fast) algorithm for {\textit{micro}} lensing studies . When we apply the algorithm to computing {\textit{weak}} and {\textit{strong}} lensing quantities in N-Body simulations, the errors are dominated by the Poisson noise caused by particle discreteness. The Poisson noise can be suppressed by smoothing out the particle distribution, which can be achieved by simply choosing an adaptive softening length in the PP calculation. We have presented a criterion to set the adaptive softening length.  Our algorithm is also applicable to cosmological simulations. We provide a \textsc{python} implementation \texttt{P3Mlens} for this algorithm.   

\end{abstract}



\section{Introduction} \label{sec:1}
Nowadays gravitational lensing becomes a powerful tool to investigate the distribution of dark matter, both  for dark matter halos and for larger scale structures \citep{2010ARA&A..48...87T,2018ARA&A..56..393M}.  With the advent of ongoing or planed large imaging and spectroscopic surveys such as LSST \citep{2019ApJ...873..111I}, DESI \citep{2019AJ....157..168D}, HSC-SSP \citep{2018PASJ...70S...4A} and PFS \citep{2014PASJ...66R...1T}, the statistical uncertainties of observational measurement for gravitational lensing will be very small, but the accuracy will be limited by remaining systematics in the observation.

Accurate theoretical modeling will be required to properly interpret the lensing observations. 
A widely used way for modeling is to perform ray-tracing computation on the basis of high resolution N-body simulations \citep{1998ApJ...494...29W,2000ApJ...530..547J,2003MNRAS.340..580T}. In nearly all the ray-tracing methods, single or multiple plane approximation is used \citep{1994A&A...287....1B,2006MNRAS.371..750H,2008MNRAS.391..435F}, where the 3D matter distribution around each plane is projected onto the plane (the 2D distribution). Thus,  the main step in the ray-tracing computation is to accurately calculate the deflection angle of light for the lens plane, which can be reduced to a physical problem of solving a 2D Poisson's equation.

In the literature, 2D Poisson's equation is usually solved using a Particle-Mesh (PM) algorithm. At first, mass of particles is assigned to a grid used for Fast Fourier Transform (FFT). Then, potential is obtained by solving Poisson's equation in Fourier space with a Green function where $\hat{G}(\bm{k})\propto1/k^2$ is often used. Differential is approximated by finite-difference to obtain the force field or the lensing parameters on the grid from the potential, and finally the lensing quantities on the grid are interpolated to the whole space. Although the Green function is correct analytically, because the calculation is done on the grid with finite interval in the PM, errors are inevitably generated due to under-sampling (alias) and anisotropies in each step from mass assignment to force interpolation. This is usually called the {\it poor man's} Poisson Solver. \cite{1981csup.book.....H} provides a method to minimize the error in the whole PM process by regarding the Green function as a free parameter and optimizing it. The force has to be softened on small scale in the PM calculation even with the optimized Green function. Thus, the PP algorithm is employed to compensate for the force softening in the PM calculation, and the force can be calculated to a very high precision at all scales. The method is usually called the Particle-Particle-Particle-Mesh (P$^3$M) algorithm.

Another source of error in ray-tracing simulations is Poisson noise. Since particle distribution in N-body simulation is just a Monte Carlo sampling of the underlining density distribution, Poisson noise is unavoidable. Smoothing is the main method to reduce the noise and recover the density distribution, which has been investigated in many studies \citep{2004A&A...423..797B,2006ApJ...652...43L,2009ChA&A..33..121Y}. Kernels of Gaussian or other shapes with an adaptive smoothing length are usually used in PM algorithm when performing mass assignment. In P$^3$M algorithm, as we will show, the smoothing can be fully incorporated in the PP calculation by transforming  point sources to shaped sources. This is another advantage of the P$^3$M algorithm to predict lensing quantities in N-Body simulations. Furthermore, we can easily adopt an adaptive smoothing length in the algorithm.    

In this paper, we present a P$^3$M algorithm for lensing studies in N-Body simulation. We study the accuracy of the computed lensing quantities, and investigate how the errors are related to the accuracy of the force calculation and to the Poisson noise of the discrete particle distribution. We give an analytical form of the optimized Green function and determine the best choice of the free parameters in PM. We also give an analytical form for the force between 2D shaped particles in real space, which is used both for the PP calculation and for the smoothing of the particle distribution. Our recommendation for the two softening (smoothing) lengths is given. Finally, we will briefly introduce our \textsc{python} implementation \texttt{P3MLens} of the 2D P$^3$M algorithm. 

This paper is organized as follows. In Section \ref{sec:2}, we briefly introduce the gravitational lensing theory and the general P$^3$M algorithm. In Section \ref{sec:3}, our realization of the 2D P$^3$M algorithm is provided. In Section \ref{sec:test}, we test the accuracy of the algorithm. The \textsc{python} implementation is briefly introduced in Section \ref{sec:p3mlens} and our conclusions are summarized in Section \ref{sec:4}. Throughout this paper, Fourier transform of a function $f$ is denoted as $\hat{f}$, and the convention of Fourier transform we use, as example for one dimension, is
\begin{equation}
    f(x)=\int_{-\infty}^{\infty}\frac{dk}{2\pi}\hat{f}(k)e^{ikx}
\end{equation}
\begin{equation}
    \hat{f}(k)=\int_{-\infty}^{\infty}dxf(x)e^{-ikx}
\end{equation}

\section{Method} \label{sec:2}
\subsection{Gravitational lensing for thin lens}
The deflection angle of light by a gravitational lens is the key quantity calculated in the ray tracing simulations. For a thin lens, according to General Relativity (GR), the deflection angle $\bm{\alpha}$ depends on the gravitational potential $\Phi$ as follows:
\begin{equation}
    \bm{\alpha} = \frac{2}{c^2}\int\bm{\nabla}_2\Phi\ dl = \frac{2}{c^2}\bm{\nabla}_2\int\Phi\ dl   \,\,,
\end{equation}
where $\bm{\nabla}_2$ is the 2D gradient operator in the lens plane. Using Poisson's equation for 3D gravitational potential, we can easily prove,
\begin{align}
    \bm{\nabla}_2^2\int\Phi\ dl &= \int\nabla^2\Phi-\frac{\partial^2\Phi}{\partial l^2}\ dl \notag\\
                   &= 4\pi G\int\rho\ dl-\frac{\partial\Phi}{\partial l}\bigg|_{\infty}^{\infty} \notag\\
                   &= 4\pi G\Sigma \,\,.
\end{align}
In the above equation, $\Sigma$ is the surface density of the lens. The second term on the {\it rhs}  vanishes due to the zero boundary condition for the potential. It's convenient to define 2D potential $\psi=\int \Phi\ dl$ and field $\bm{E}=-\bm{\nabla}_2\psi$. We have $\bm{\alpha} = -\frac{2}{c^2}\bm{E}$. Thus, the goal of ray-tracing simulations is to solve the 2D Poisson's equation from the known matter distribution of an N-body simulation, 
\begin{equation}
\bm{\nabla}_2^2\psi=4\pi G\Sigma
\end{equation} 
It's easy to derive the 2D force field for a point source:
\begin{equation}
    \bm{E} = -\frac{2Gm\bm{r}}{r^2}\,\,,
\end{equation}
where m is the mass of the point source. 

For a single thin lens, gravitational lensing parameters like shear ($\gamma_1$, $\gamma_2$), convergence ($\kappa$) and magnification ($\mu$) can be calculated if $\bm{\alpha}$ is known in the lens plane. 
\begin{equation}
    \gamma_1 = \frac{D_lD_{ls}}{2D_s}(\frac{\partial\alpha_x}{\partial x}-\frac{\partial\alpha_y}{\partial y})
\end{equation}
\begin{equation}
    \gamma_2 = \frac{D_lD_{ls}}{2D_s}(\frac{\partial\alpha_x}{\partial y}+\frac{\partial\alpha_y}{\partial x})
\end{equation}
\begin{equation}
    \bm{\gamma} = \gamma_1+i\gamma_2
\end{equation}
\begin{equation}
    \kappa = \frac{D_lD_{ls}}{2D_s}(\frac{\partial\alpha_x}{\partial x}+\frac{\partial\alpha_y}{\partial y})
\end{equation}
\begin{equation}
    \mu = \frac{1}{(1-\kappa)^2-|\gamma|^2}
\end{equation}
where $D_l$, $D_s$ and $D_{ls}$ are the angular diameter distances from the observer to the lens, the observer to the source and from the lens to the source.
\subsection{General multi-dimensional P$^{3}$M algorithm}
P$^{3}$M algorithms are a class of hybrid algorithms developed decades ago to simulate plasma systems using particles. They have been used to simulate the evolution of large-scale structures in the Universe (e.g. \cite{1985ApJS...57..241E,1991ApJ...368L..23C,1994A&A...284..703J}). The short-range force on a particle is computed by directly summing up particle-particle (PP) pair force, and the smoothly varying long-range force is approximated by the particle-mesh (PM) force calculation. In this subsection, we briefly introduce the main idea of the P$^{3}$M algorithm and we recommend \cite{1981csup.book.....H} for details. The main steps of the P${^3}$M algorithm include: PM (mass assignment, potential solving, potential difference and force interpolation) and PP (force splitting and summation) calculations.

\subsubsection{Mass assignment}
To take advantage of the Fast Fourier Transform (FFT) algorithm, the density field is assigned to a regular grid. Assignment schemes commonly adopted are Nearest Grid Point (NGP), Cloud In Cell (CIC), Triangular Shape Cloud (TSC) and Piecewise Cubic Spline (PCS), which corresponds to the zeroth, first, second and third order piecewise polynomial functions $W^{(p)}$ with $p=0$,1,2,3 respectively. For each of the schemes, we have a one-dimensional function: 
\begin{equation}
    W^{(0)}(x) = 
    \begin{cases}
    1 & |x|<\frac{1}{2}\\
    0 & otherwise
    \end{cases}
\end{equation}
\begin{equation}
    W^{(1)}(x) = 
    \begin{cases}
    1-|x| & |x|< 1\\
    0 & otherwise
    \end{cases}
\end{equation}
\begin{equation}
    W^{(2)}(x) = 
    \begin{cases}
    \frac{3}{4}-x^2 & |x|< \frac{1}{2}\\
    \frac{1}{2}(\frac{3}{2}-|x|)^2 & \frac{1}{2}\leq|x|<\frac{3}{2}\\
    0 & otherwise
    \end{cases}
\end{equation}
\begin{equation}
    W^{(3)}(x) = 
    \begin{cases}
    \frac{1}{6}(4-6x^2+3|x|^3) & 0\leq|x|< 1\\
    \frac{1}{6}(2-|x|)^3 & 1\leq|x|<2\\
    0 & otherwise
    \end{cases}
\end{equation}
The mass assignment weight function is given by $W(\bm{x}) = \prod\limits_{j}^dW^{(p)}(x^j/H)$, where $d$ is the dimension and $H$ is the linear size of the grid cell.

For a system with the number density of particles:
\begin{equation}
    n(\bm{x}) = \sum\limits_{i=1}^{N_p}\delta(\bm{x}-\bm{x_i})
\end{equation}
where $\bm{x_i}$ is the position of particle $i$, $N_P$ is the number of particles and $\delta(\bm{x})$ is the Dirac delta function. The mass density after assignment can be described by:
\begin{equation}
    \rho(\bm{x}) = \frac{m}{V_g}\int n(\bm{x'})W(\bm{x}-\bm{x'})d\bm{x'}
\end{equation}
where $m$ is the particle mass and $V_g$ is the volume of the grid. The mass density grid used for FFT can be generated by a sampling process described by the sampling function:
\begin{equation}
    \text{III}(\bm{x}) = \sum\limits_{\bm{n}}\delta(\bm{x}-\bm{n})
\end{equation}
\begin{equation}
    \rho'(\bm{x}) = \text{III}(\frac{\bm{x}}{H})\rho(\bm{x})
\end{equation}
with $\bm{n}$ an integer vector for denoting the coordinates of the grid points. Thus, the whole mass assignment process can be expressed as:
\begin{equation}
    \rho'(\bm{x}) = \frac{m}{V_g}\text{III}(\frac{\bm{x}}{H})W(\bm{x})\bm{*}n(\bm{x}) \,,\label{eq:18}
\end{equation}
where and below a convolution is denoted by $\bm{*}$. 
\subsubsection{Potential solving, finite difference and force interpolation}
The Green function $G(\bm{x})$ describes the potential from a unit mass point source. The solution of Poisson's equation can be written as:
\begin{equation}
    \psi(\bm{x}) = G(\bm{x})\bm{*}\rho'(\bm{x})\label{eq:19}
\end{equation}
The equation is usually calculated in Fourier space ($\bm{k}$) in which the convolution is transformed to a simple multiplication product. In the literature, there are many choices for the Green function, e.g.,  the "poor man's" Green function $\hat{G}(\bm{k})\propto1/k^2$. However, here we will treat it as a free parameter which will be optimized after we take into account of all the errors in the PM calculation.

The force field is calculated from the potential by $\bm{E} = -\bm{\nabla}\psi$. However, the differential is approximated by a difference in PM:
\begin{equation}
    \bm{E}(\bm{x}) = -\bm{D}(\bm{x})\bm{*}\psi(\bm{x})\label{eq:20}
\end{equation}
 The first order approximation for the differential is the two-point finite-difference,
\begin{align}
    D(x) &= \frac{\delta(x+H)-\delta(x-H)}{2H}
\end{align}
\begin{equation}
    \bm{D}(\bm{x}) = \sum_{j=1}^dD(x^j)\bm{\hat{x}}^j
\end{equation}
The error of the difference is at the order of $O(H^2)$. And the next order approximation is the  four-point finite-difference,
\begin{align}
    D(x) &= \beta\frac{\delta(x+H)-\delta(x-H)}{2H} \notag\\
         &+(1-\beta)\frac{\delta(x+2H)-\delta(x-2H)}{4H}
\end{align}
Setting $\beta=4/3$ leads to a cancellation of $O(H^2)$ term and the accuracy of order $O(H^4)$ can be achieved. Higher order approximations can be more accurate,  but they are also more time-consuming in the calculation since more points are used for the difference.
The force field grid $\bm{E}'$ calculated by PM is also a sampling of the underlining force field $\bm{E}$:
\begin{equation}
    \bm{E}'(\bm{x}) = -\text{III}(\frac{\bm{x}}{H})\bm{D}(\bm{x})\bm{*}\psi(\bm{x})  \,\,.\label{eq:23}
\end{equation}

Finally, the force field should be interpolated to the whole space. We use the same interpolation function for the force as we did for the mass assignment,
\begin{equation}
    \bm{F}(\bm{x})=\frac{m}{V_g}W(\bm{x})\bm{*}\bm{E}'(\bm{x})\,\,.\label{eq:24}
\end{equation}
When the two interpolation functions are taken the same, the momentum of the system is conserved.

\subsubsection{Optimized Green function}
Though the Green function ($\hat{G}(\bm{k})\propto1/k^2$) is accurate in {\it poor man's} Poisson Solver, errors are generated in every step of the PM calculation. For example, in the mass assignment, error called alias is generated during sampling if $\hat{\rho}(\bm{k})$ is non-zero beyond the Nyquist frequency:
\begin{equation}
    \hat{\rho}'(\bm{k}) = \sum\limits_{\bm{n}=-\infty}^{+\infty}\hat{\rho}(\bm{k}-\bm{n}k_g)\label{eq:27}
\end{equation}
with $k_g=2\pi/H$. Using a finite-difference to replace differential also generates error, and so does the interpolation. Thus, the Green function is better regarded as a free parameter to be optimized to compensate for all the errors arising in the PM calculation.

The Green function is optimized to minimize the average force error between two particles. The force on a unit mass particle at $\bm{x}_2$ from another unit mass particle at $\bm{x}_1$ is given by:
\begin{equation}
    \bm{F}(\bm{x}_2) = \int\frac{d\bm{k}}{(2\pi)^d}\bm{\hat{F}}(\bm{k})e^{i\bm{k}\cdot\bm{x}_2}
\end{equation}
From equations \ref{eq:18}-\ref{eq:20} and \ref{eq:23}-\ref{eq:27}, we get
\begin{align}
    \bm{\hat{F}}(\bm{k}) &= \frac{1}{V_g}\hat{W}\hat{\bm{E}}'\notag\\
                         &= -\frac{1}{V_g}\hat{W}\hat{\bm{D}}\hat{\psi}\notag\\
                         &= -\frac{1}{V_g}\hat{W}\hat{\bm{D}}\hat{G}\hat{\rho}'\notag\\
                         &= -\frac{1}{V_g^2}\hat{W}\hat{\bm{D}}\hat{G}\sum\limits_{\bm{n}}\hat{W}(\bm{k}-\bm{n}k_g)e^{-i(\bm{k}-\bm{n}k_g)\cdot\bm{x}_1}\label{eq:26}
\end{align}
The last step uses the Fourier transform of the $\delta$ function that ${\delta}(\bm{k})=e^{-i\bm{k}\cdot\bm{x}_1}$. Thus, the force between two particles in a periodic system becomes:
\begin{align}
    \bm{F}(\bm{x}_2; \bm{x}_1)=\frac{1}{V_b}\sum\limits_{\bm{l}}\frac{\hat{W}}{V_g}\bm{\hat{D}}\hat{G}\sum\limits_{\bm{n}}\frac{\hat{W}(\bm{k}_{\bm{n}})}{V_g}e^{-i\bm{k}_{\bm_n}\cdot\bm{x}_1}e^{-i\bm{k}\cdot\bm{x}_2}
\end{align}
Where $\bm{l}$ is the wavenumber of the Fourier series for the periodic system, $V_b$ is the box volume and $\bm{k}_{\bm{n}}=\bm{k}-\bm{n}k_g$.

The displacement-averaged total squared deviation of the PM calculated force $F(\bm{x} = \bm{x}_2-\bm{x}_1; \bm{x}_1)$ from the reference interparticle force $\bm{R}(\bm{x})$ is defined as:
\begin{equation}
    Q=\frac{1}{V_g}\int_{V_g}d\bm{x}_1\int_{V_b}d\bm{x}|\bm{F}(\bm{x};\bm{x}_1)-\bm{R}(\bm{x})|^2
\end{equation}
Using Fourier transform, $Q$ can be written as:
\begin{align}
    Q &= \frac{1}{V_b}\sum\limits_{\bm{l}}\bigg\{|\hat{\bm{D}}|^2\hat{G}^2\bigg[\sum\limits_{\bm{n}}\hat{U}(\bm{k}_{\bm{n}})\bigg]^2\notag\\
    &-2\hat{G}\bm{\hat{D}}\cdot\sum\limits_{\bm{n}}\hat{\bm{R}}^{\bm{*}}(\bm{k}_{\bm{n}})\hat{U}^2+\sum\limits_{\bm{n}}|\hat{\bm{R}}|^2\bigg\}\label{eq:Q}
\end{align}
where $\hat{U} = \hat{W}/V_g$. Minimizing $Q$ with respect to $\hat{G}$ gives the optimized Green function:
\begin{equation}
    \hat{G}(k) = \frac{\bm{\hat{D}}(\bm{k})\cdot\sum\limits_{\bm{n}}\hat{U}^2(\bm{k_n})\bm{\hat{R}}^{\ast}(\bm{k_n})}{|\bm{\hat{D}}(\bm{k})|^2\left[\sum\limits_{\bm{n}}\hat{U}^2(\bm{k_n})\right]^2}
\end{equation}
\subsubsection{PP force calculation}
Due to the very nature of the PM calculation, the PM force is both softened and anisotropic on small scale (smaller than a few $H$) compared to the reference force between two point particles. The PP algorithm is usually used to accurately calculate the short-range force, in order to compensate for the PM softening. To suppress the error from the anisotropy and to smoothly transit from the PM force to the PP force, a shape $S(r)$ is assigned to each particle for the PM, reflected in the softened $\bm{R}$:
\begin{equation}
    \hat{\bm{R}} = \hat{\bm{R}}_p\hat{S}^2
\end{equation}
where $\hat{\bm{R}}_p$ is the Fourier transform of the force between two point sources. The shapes with polynomial forms are usual choices for convenient analytical calculation, such as:
\begin{equation}
    S_1(r;a) = 
    \begin{cases}
    32/\pi a^4\big(a^2/4-r^2\big)& r<a/2\\
    0 & otherwise
    \end{cases}\label{eq:s1}
\end{equation}
\begin{equation}
    S_2(r;a) = 
    \begin{cases}
    24/\pi a^3\big(a/2-r\big)& r<a/2\\
    0 & otherwise
    \end{cases}\label{eq:s2}
\end{equation}
in the 2D case. The smoothing length is controlled by $a=a_{pm}$, which also influences the volume to perform PP. Larger $a_{pm}$ can better suppress the total error at the cost of many more particle pairs in PP ($O(a_{pm}^d)$). The force used for PP is simply $\bm{F}_{pp}=\bm{R}_{tot}-\bm{R}_{pm}$, where $\bm{R}_{tot}$ and $\bm{R}_{pm}$ are the reference total force and PM force respectively. 

\section{2D P$^3$M algorithm for gravitational lensing}\label{sec:3}
First of all, the analytical form of the optimized Green function should be obtained. The analytic forms of $\hat{\bm{D}}$ and $\hat{\bm{U}}$ are provided in \cite{1981csup.book.....H}:
\begin{align}
    &\hat{D}_x(\bm{k}) = i\beta\frac{\sin{(k_xH)}}{H}+i(1-\beta)\frac{\sin{(2k_xH)}}{2H}\notag\\
    &\hat{D}_y(\bm{k}) = i\beta\frac{\sin{(k_yH)}}{H}+i(1-\beta)\frac{\sin{(2k_yH)}}{2H}\label{eq:35}
\end{align}
\begin{equation}
    \hat{U}(\bm{k}) = \left(\frac{\sin{(\frac{k_xH}{2})}\sin{(\frac{k_yH}{2}})}{\frac{k_xH}{2}\frac{k_yH}{2}}\right)^{p+1}\label{eq:36}
\end{equation}
where $p$ is the order of the piecewise polynomial function used for mass assignment and force interpolation. 

Obtaining $\bm{\hat{R}}$ in 2D is more complicated than in 3D. We first calculate $\hat{\bm{R}}_p$ and then consider $\hat{S}$ for a selected particle shape. For unit mass, $\bm{R}_{p}(\bm{r})=2G\bm{r}/r^2$, the 2D Fourier transform of the x component is: 
\begin{align}
    \hat{R}_{px}(\bm{k})&=2G\int_{\infty}^{\infty}\int_{\infty}^{\infty}\frac{x}{x^2+y^2}e^{-(ik_xx+ik_yy)}dxdy\notag\\
                &=-2\pi Gi\int_{\infty}^{\infty}sgn(k_x)e^{-|k_xy|-ik_yy}dy\notag\\
                &=-4\pi Gi\frac{sgn(k_x)}{|k_x|}\frac{k_x^2}{k_x^2+k_y^2}\notag\\
                &=-4\pi G\frac{ik_x}{k_x^2+k_y^2} 
\end{align}
Where $sgn$ is the sign function. Similarly we have:
\begin{align}
    &\hat{R}_{py}(\bm{k}) = -4\pi G\frac{ik_y}{k_x^2+k_y^2}\notag\\
    &\hat{\bm{R}}_p(\bm{k}) = -4\pi G\frac{i\bm{k}}{k^2}\label{eq:38}
\end{align}
Two forms of particle shape, Eqs. \ref{eq:s1} and \ref{eq:s2}, are considered and compared here. $S_2$ is widely used in 3D simulations, which is however not preferred in the 2D case as will be shown. The Fourier transform of $S_1$ is,
\begin{align}
    \hat{S}_1(k;a)=\int_0^{2\pi}e^{-ikr\cos{\theta}}d\theta\int_0^{\frac{a}{2}}\frac{32}{\pi a^4}(\frac{a^2}{4}-r^2)rdr\,\,,
\end{align}
where $\theta$ is the angle between $\bm{r}$ and $\bm{k}$. Using Jacobi–Anger expansion,
\begin{align}
    e^{iz\cos{\theta}} = J_0(z) + 2\sum\limits_{n=1}^{+\infty}i^nJ_n(z)\cos{(n\theta)}\,\,,
\end{align}
where $J_n$ is the $n$-th order Bessel functions of the first kind, we get
\begin{align}
    \hat{S}_1(k;a)&=\frac{64}{a^4}\int_0^{\frac{a}{2}}(\frac{a^2}{4}r-r^3)J_0(kr)dr\notag\\
    &=\frac{128}{k^3a^3}J_1(\frac{ka}{2})-\frac{32}{k^2a^2}J_0(\frac{ka}{2})\label{eq:43}
\end{align}
Similarly we can derive $\hat{S}_2$,
\begin{align}
    \hat{S}_2(k;a) = \frac{12}{k^2a^2}[J_1(\frac{ka}{2}){\rm{\bm{H}}}_0(\frac{ka}{2})-J_0(\frac{ka}{2}){\rm{\bm{H}}}_1(\frac{ka}{2})]\label{eq:44} \,\,,
\end{align}
where ${\rm{\bm{H}}_n}$ is the $n$-th Struve function.
\begin{figure}
    \centering
    \includegraphics[scale=0.45]{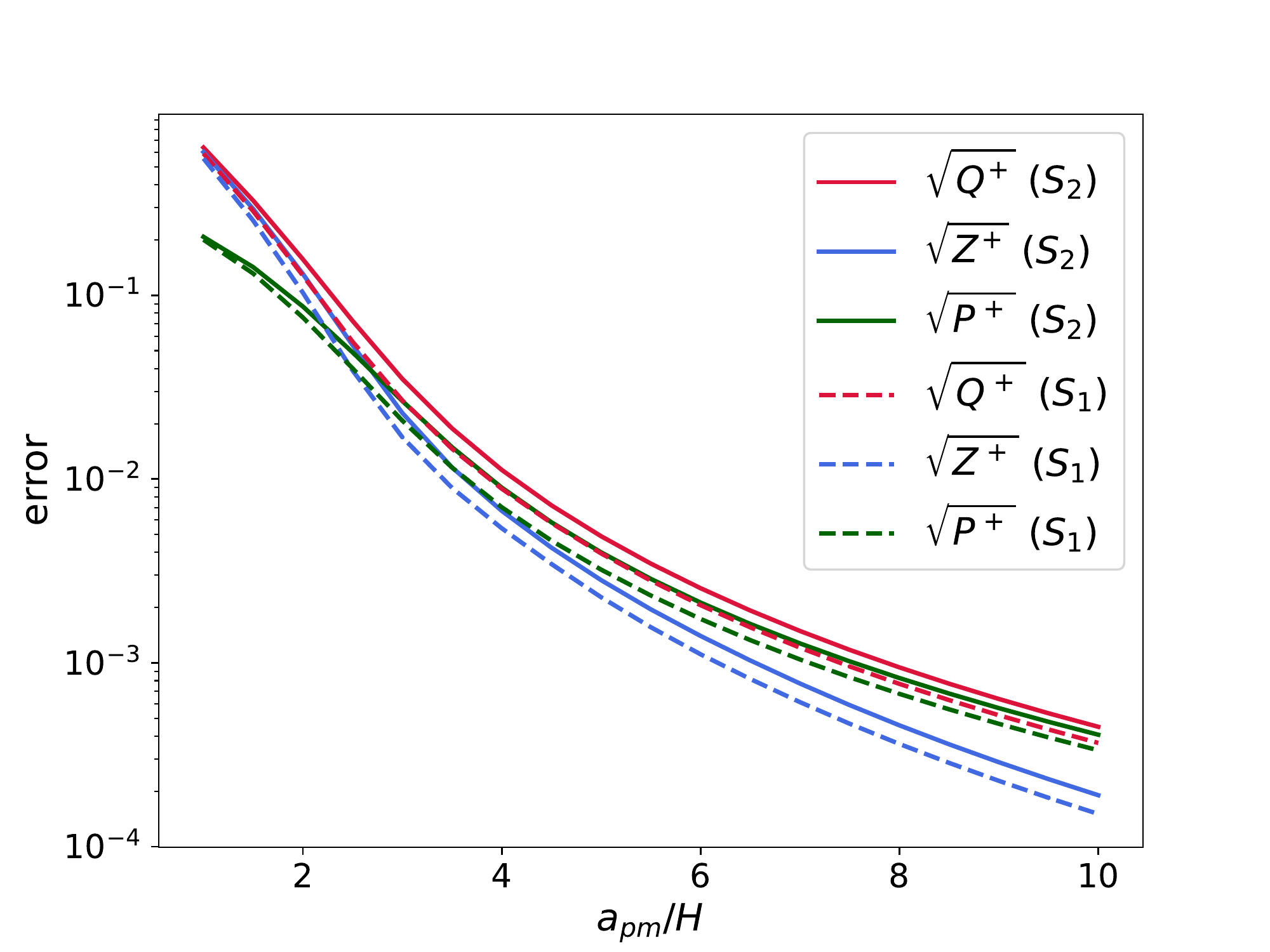}
    \caption{The dependence of $\sqrt{Q^+}$, $\sqrt{Z^+}$ and $\sqrt{P^+}$ on $a_{pm}$. The solid lines show the results for particle shape $S_2$ and dashed ones for $S_1$.}
    \label{fig:1}
\end{figure}

The analytic form of $\hat{G}$ can be derived with equations \ref{eq:35}-\ref{eq:36},\ref{eq:38},\ref{eq:43} and \ref{eq:44}. The summation $\sum\limits_{\bm{n}}$ can be done within $|\bm{n}|< 2$, for very fast decay of $\hat{U}^2$ and $\hat{\bm{R}}$ with respect to $|\bm{k}|$. 

Free parameters $p$, $a_{pm}$ and the shape form remain to be determined. The main improvement with higher order mass assignment $p$ is to suppress the anisotropy \citep{1985ApJS...57..241E}. Thus, in the later part of this paper, we set $p=2$ by default with the TSC mass assignment scheme. As we will show, with this scheme, we can achieve a sufficiently good isotropy with a reasonable choice of  $a_{pm}$.

The error $Q$ (equation \ref{eq:Q}) in PM with different choices of particle shape and  $a_{pm}$ can be calculated for the optimized Green function $\hat{G}$. To investigate the sources of the error, $Q$ can be split into two parts,
\begin{align}
    Z &= \int d\bm{x}|\langle \bm{F}(\bm{x};\bm{x}_1)\rangle-\bm{R}(\bm{x})|^2\notag\\
    &= \frac{1}{V_b}\sum\limits_{\bm{l}}\bigg\{|\hat{\bm{D}}|^2\hat{G}^2\bigg[\sum\limits_{\bm{n}}\hat{U}^4\bigg]\notag\\
    &-2\hat{G}\hat{\bm{D}}\cdot\bigg[\sum\limits_{\bm{n}}\hat{\bm{R}}^{\bm{*}}\hat{U}^2\bigg]+\sum\limits_{\bm{n}}|\hat{\bm{R}}|^2\bigg\}\,\,,
\end{align}
and
\begin{align}
    P &= \frac{1}{V_g}\int_{V_g}d\bm{x}_1\int_{V_b}d\bm{x}|\bm{F}(\bm{x};\bm{x}_1)-\langle \bm{F}(\bm{x};\bm{x}_1)\rangle|^2\notag\\
    &= \frac{1}{V_b}\sum\limits_{\bm{l}}|\hat{\bm{D}}|^2\hat{G}^2\bigg\{\bigg[\sum\limits_{\bm{n}}\hat{U}^2\bigg]^2-\sum\limits_{\bm{n}}\hat{U}^4\bigg\}\,\,.
\end{align}

In the above equations, 
$Z$ is the total squared deviation of the displacement-averaged mesh force from the reference force, which describes the deviation of the mean force at separation $r$ from the reference force,  and $P$ is the total squared deviation of the mesh force from its displacement average, which reflects the anisotropy of the PM force. By definition, we have $Q=Z+P$. To construct the dimensionless error estimator, we define
\begin{align}
    &Q^{+}=\frac{Q}{(\frac{2G}{H})^2(\pi H^2)}\\
    &Z^{+}=\frac{Z}{(\frac{2G}{H})^2(\pi H^2)}\\
    &P^{+}=\frac{P}{(\frac{2G}{H})^2(\pi H^2)} \,\,.
\end{align}

\begin{figure}
    \centering
    \includegraphics[scale=0.43]{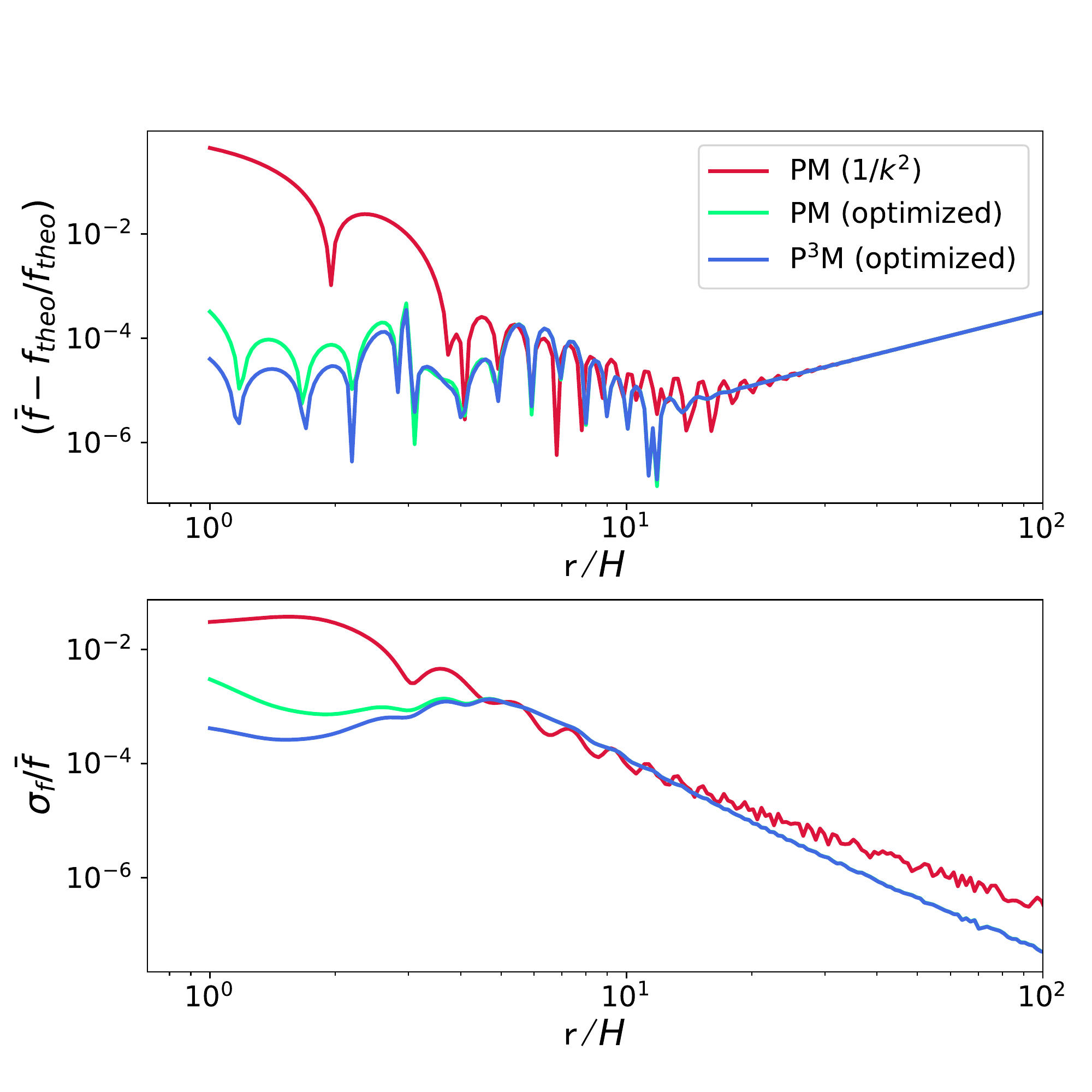}
    \caption{Top: The mean deviation of the computed force from the theoretical prediction at distance $r$. Bottom: Anisotropy, defined as the $rms$ deviation of the force field ($\sigma_{f}$) divided by the mean force field ($\bar{f}$), at distance $r$ from a point source. The blue lines show the results for the P$^3$M algorithm and the red ones for the conventional PM algorithm for a point source ($2G/r$). The green lines shows the results for the PM algorithm with the optimized Green function for a soften source ($R(r;a)$).}
    \label{fig:2}
\end{figure}

The dependence of $\sqrt{Q^+}$, $\sqrt{Z^+}$ and $\sqrt{P^+}$ on $a_{pm}$ is shown in Figure \ref{fig:1}. Results for the two different particle shapes are also presented. In general, $S_1$ performs better than $S_2$. And considering that the calculation of $J_n$ is much faster than ${\rm{\bm{H}_n}}$, we prefer $S_1$ in our 2D lensing calculation.  As expected, with increasing $a_{pm}$, the error is decreasing. At small $a_{pm}$, the error is mainly from the deviation of the mean force from the reference force ($Z$). Gradually, the anisotropy ($P$) becomes a dominant part of the error as $a_{pm}$ increases. The decrease of the error is much slower once $a_{pm}$ reaches $6H$. Considering the computational cost increases with $a_{pm}$ in the PP calculation,  we recommend $a_{pm}\approx6H$. Moreover, using higher order finite-difference approximations may also reduce the error, which we don't investigate further here, as the PM force is accurate enough (better than 0.1 percent on all scales, cf. Fig.\ref{fig:2}) with the above parameters.

To perform PP on the basis of PM, an analytic form of $\bm{F}_{pp}(\bm{r})$, determined by $\bm{R}_{tot}$ and $\bm{R}_{pm}$, is needed. $\bm{R}_{tot}$ should also be smoothed to recover the underlining matter distribution from the Monte Carlo sampling particles. Hence, real space force $\bm{R}(\bm{r};a)$ between two particles of unit mass with shape $S_1$ for any $a$ should be known. Actually, knowing the magnitude of the force $R(r;a)$ at distance $r$ is enough since $R$ doesn't depends on the direction of $\bm{r}$ and the direction of $\bm{R}$ can be easily calculated from the particle positions.
The Fourier transform of the $x$ component of $\bm{R}(\bm{r};a)$ is:
\begin{equation}
    \hat{R}_{x}(k_x,k_y) = -4\pi G\frac{ik_x\hat{S}_1^2(k;a)}{k^2}\,\,.
\end{equation}
Transforming it to real space,
\begin{align}
    R_{x}(x,y) = -\frac{4\pi G}{(2\pi)^2}\int_{\infty}^{\infty}\int_{\infty}^{\infty}\frac{ik_x\hat{S}_1^2(k)}{k^2}e^{ik_xx+ik_yy}dk_xdk_y \,\,,
\end{align}
we can get $R(r;a)$ from $R_{x}(x,y)$ by setting $x=r$ and $y=0$,
\begin{align}
    R(r;a) & = R_{x}(r,0) \notag\\
               & = -\frac{G}{\pi}\int_{\infty}^{\infty}\int_{\infty}^{\infty}\frac{ik_x\hat{S}_1^2(k)}{k^2}e^{ik_xr}dk_xdk_y\notag\\
               & = -\frac{G}{\pi}\int_{0}^{\infty}\int_{0}^{2\pi}i \cos{\theta}\hat{S}_1^2(k)e^{ikr\cos{\theta}}dkd\theta \,\,.
\end{align}
\begin{figure}
    \centering
    \includegraphics[scale=0.57]{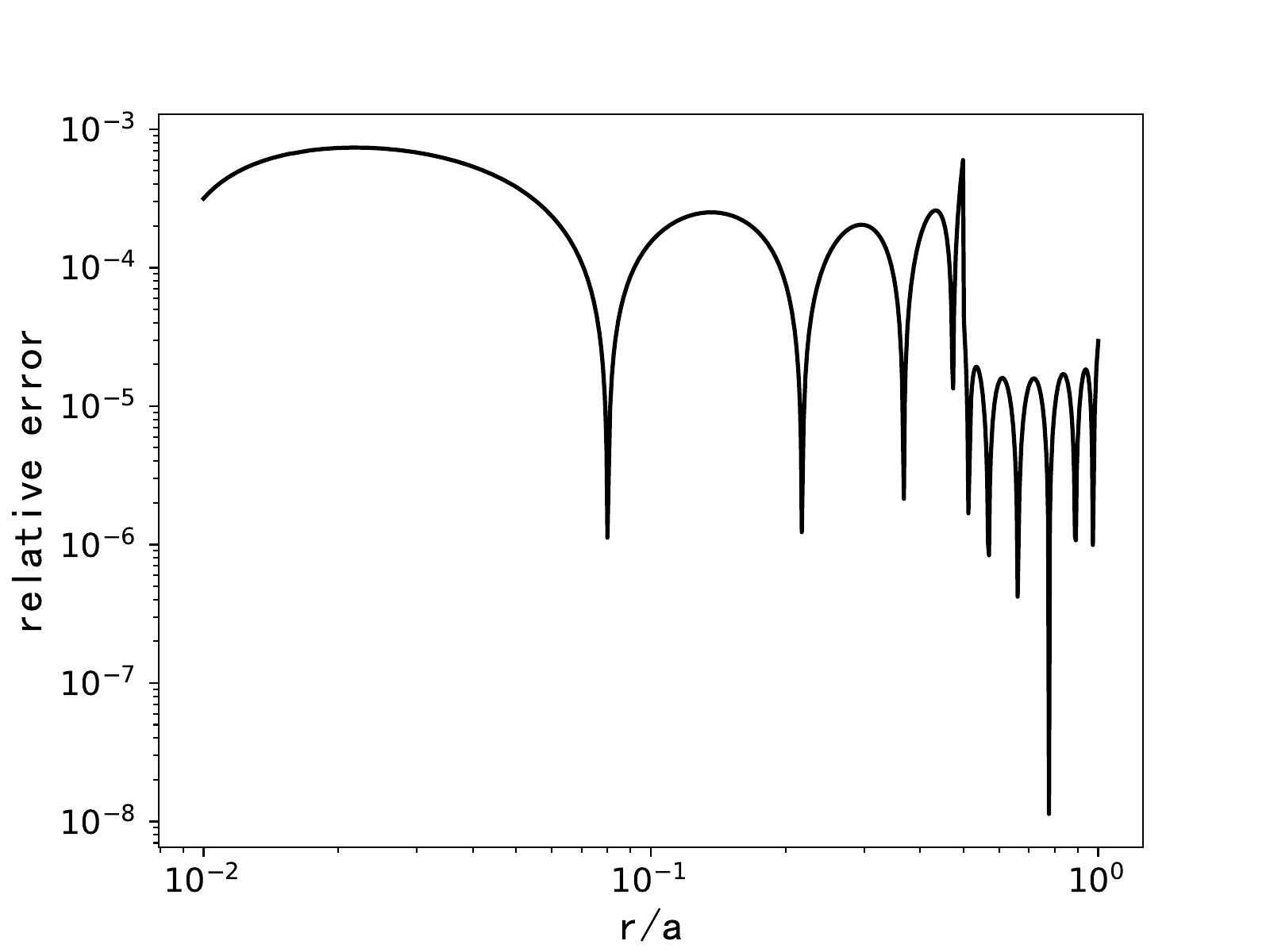}
    \caption{Relative error of the fitting formula (Equation \ref{eq:fit}) of the real space force $R(r;a)$.}
    \label{fig:fit}
\end{figure}
Using Jacobi–Anger expansion, and integrating $\theta$ component, we get
\begin{align}
    &\int_0^{2\pi}\cos{\theta}e^{ikr\cos{\theta}}d\theta \notag\\
    & = 2\sum\limits_{n=1}^{+\infty}\int_0^{2\pi}i^nJ_n(kr)\cos{(n\theta)}\cos{\theta}d\theta \,\,.
\end{align}
Only the $n = 1$ component survives in $\int_0^{2\pi}\cos{(n\theta)}\cos{\theta}d\theta$, thus
\begin{align}
    \int_0^{2\pi}\cos{\theta}e^{ikr\cos{\theta}}d\theta &= 2\int_0^{2\pi}iJ_1(kr)\cos^2{\theta}d\theta \notag\\
    &= 2\pi iJ_1(kr) \,\,.
\end{align}
Finally we have
\begin{align}
    R(r;a) = 2G\int_0^{\infty}J_1(kr)\hat{S}_1^2(k;a)dk \,\,.
\end{align}
The integration is hard to calculate analytically, so we provide an accurate ($<10^{-3}$, Figure \ref{fig:fit}) fitting formula for fast calculation:
\begin{equation}
    2G
    \begin{cases}
    \frac{1}{a}(0.743081\xi^4-1.832992\xi^3-0.057116\xi^2\\+2.672707\xi-0.000082) & 0\leq \xi < 1\\
    \frac{1}{a}(-0.524104\xi^4+4.081230\xi^3-11.856245\xi^2\\+15.107021\xi-6.823192+1.539967/\xi) & 1\leq \xi < 2\\
    1/r & 2<\xi
    \end{cases}\label{eq:fit}
\end{equation}
where $\xi=2r/a$. Thus, 
\begin{equation}
    \bm{F}_{pp}(\bm{r})=\bm{R}(\bm{r};a_{pp})-\bm{R}(\bm{r};a_{pm})
\end{equation}

The PP calculation for each point of interest should be performed within the space range $max(a_{pm},a_{pp})$ since $\bm{F}_{pp}$ vanishes beyond, so the computational cost highly depends on the the choice of these two parameters. We have discussed the choice of $a_{pm}$ already, and we will determine $a_{pp}$ in next section.

The purpose of this paper is to accurately calculate the 2D potential field $\bm{E}$, which is proportional to the deflection angle $\bm{\alpha}$. Although we only discussed about the calculation of $\bm{R}_{tot}$, it can be converted to $\bm{E}$ by simply dividing it by the particle mass.
\section{Test}\label{sec:test}
In this section, we test the accuracy of our algorithm and determine the smoothing length for gravitational lensing studies in N-body simulations.
\subsection{Point source}
We first calculate the 2D force field from a point source with our algorithm, and compare the performance with that of the {\it poor man's} PM Poisson solver ($G(k) \propto 1/k^2$) . We generate particles with different positions in the same grid cell, and calculate force field for different directions at a particular $r$. Errors of our algorithm and of the {\it poor man's} PM are shown in Figure \ref{fig:2}. The test is done with $H=1\ kpc/h$, $a_{pm}=6H$, $a_{pp}=0$, particle mass $m=5\times10^8M_{\odot}/h$ and box size $10000\ kpc/h$. The large box is used to reduce the effect of the periodic images. The mean deviation from the theoretical force field and the standard deviation in different directions of our algorithm are both two orders of magnitude smaller than the PM at $r$ smaller than $a_{pm}$, which can be attributed to our optimized PM and PP calculations. At large $r$, the mean deviation of the two algorithms is nearly the same but the standard deviation is slightly smaller in our P$^3$M algorithm due to optimized $\hat{G}$. Overall, the force between the two particles is calculated with the error smaller than $10^{-3}$ in our P$^3$M algorithm at all scales. At small $r$, the PM algorithm with optimized Green function also performs better than the conventional one (green lines). However, in P$^3$M algorithm, since the short-range force is dominated by PP calculation, the improvement in accuracy at small $r$ is mainly due to PP.

The precise calculation of forces between point sources at all $r$ makes P$^3$M an {\textit{ideal and elegant}} algorithm for micro lensing simulations. The calculation is much faster than the pure PP algorithm and the accuracy is guaranteed.
\subsection{Halo with NFW profile}
As shown in the previous subsection, we can achieve a high precision in the force calculation. However, dark matter distribution is represented by particles in N-body simulations, which inevitably leads to a Poisson noise in the lensing quantities.  To suppress the Poisson noise, one needs to smooth out the density distribution, which can be done in our algorithm by just choosing a proper softening length $a_{pp}$ in the PP calculation. The softening length needs to be large enough to smooth the density field but be small enough to maintain the underlining density distribution. We adopt an adaptive softening length $a_{pp}$ for each point of interest as the distance to the $N_{nb}$-th nearest neighbor particle. We calculate lensing quantities $\kappa$, $\gamma$ and $\mu$ for NFW \citep{1997ApJ...490..493N} halos, and compare the results with the theoretical prediction. From the comparison, we test the performance for different $N_{nb}$ and try to give a recommendation for the choice of $N_{nb}$. 
  
 We test our algorithm with an NFW halo of the virial mass $M_{vir}=10^{14.0}M_{\odot}/h$ and the virial radius $R_{vir}=709\ kpc/h$ ($z=0.5$). We use the mass-concentration relation from \cite{2014MNRAS.441.3359D}. Analytical forms of the lensing quantities for an NFW halo have been provided by \cite{2003MNRAS.340..580T} and \cite{2003MNRAS.344..857T}. We generate Monte Carlo particles for such a halo with particle mass of $5\times10^{8}M_{\odot}/h$, which is the typical mass resolution ($10^8\sim10^9M_{\odot}/h$) for cosmological simulations. The calculation is done with $H=1\ kpc/h$, $a_{pm}=6H$ and box size $6R_{vir}$. Redshift is set to be 1.0 and 0.5 for the source and lens planes respectively. 
\begin{figure*}[h]
     \centering
     \includegraphics[scale=0.6]{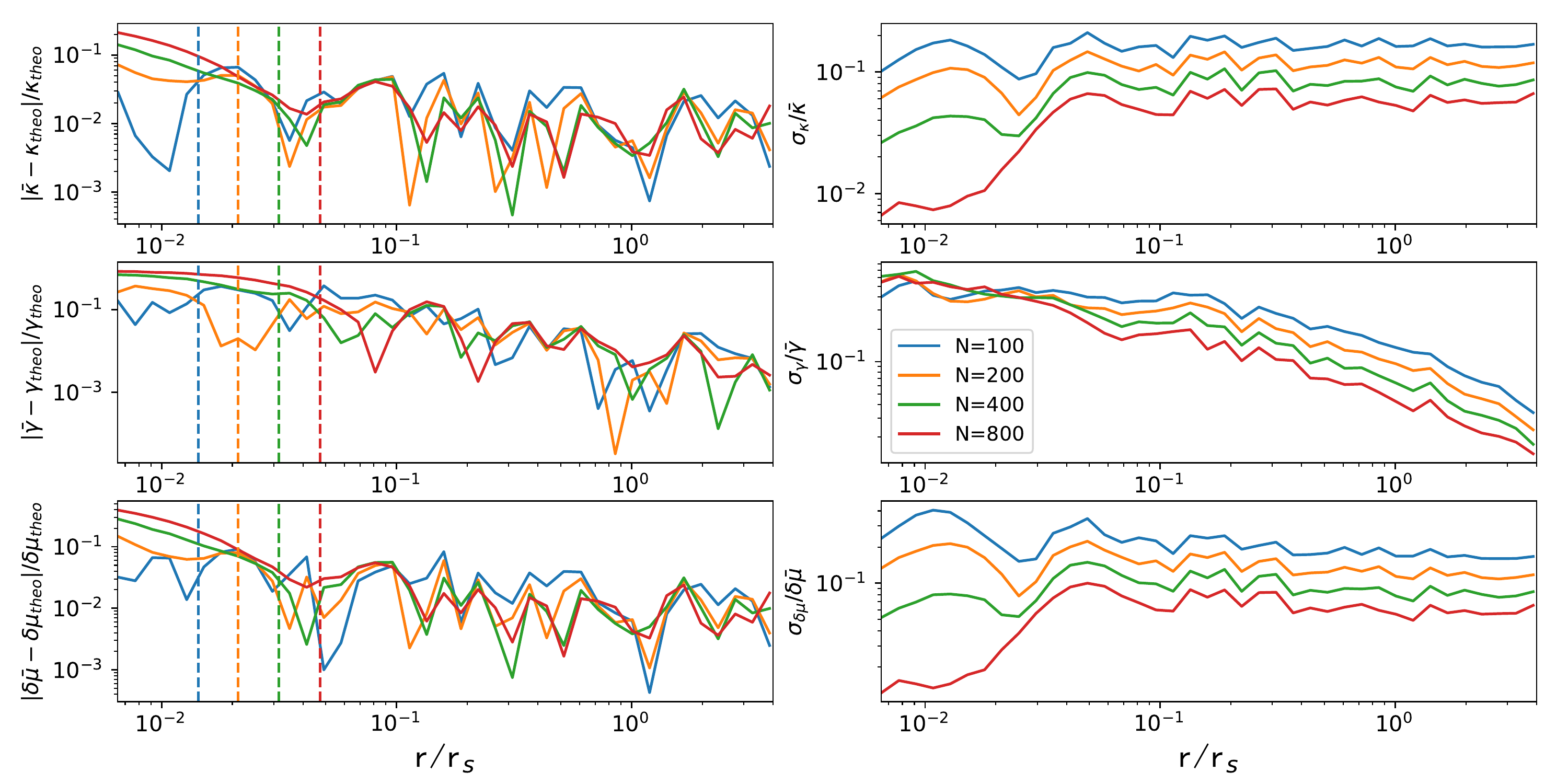}
     \caption{The relative deviation of the mean value from the theoretical prediction (left) and the standard deviation normalized by the mean value (right) for $\kappa$ (top), $\gamma$ (middle) and $\delta\mu$ (bottom) as function of radius. The results for different smoothing lengths, denoted by $N_{nb}$, are shown in different colors. For each $N_{nb}$, a vertical dashed line is drawn to indicate the smallest smoothing length $a_{min}$ around the halo center.}
     \label{fig:3}
 \end{figure*}
 \begin{figure*}
     \centering
     \includegraphics[scale=0.6]{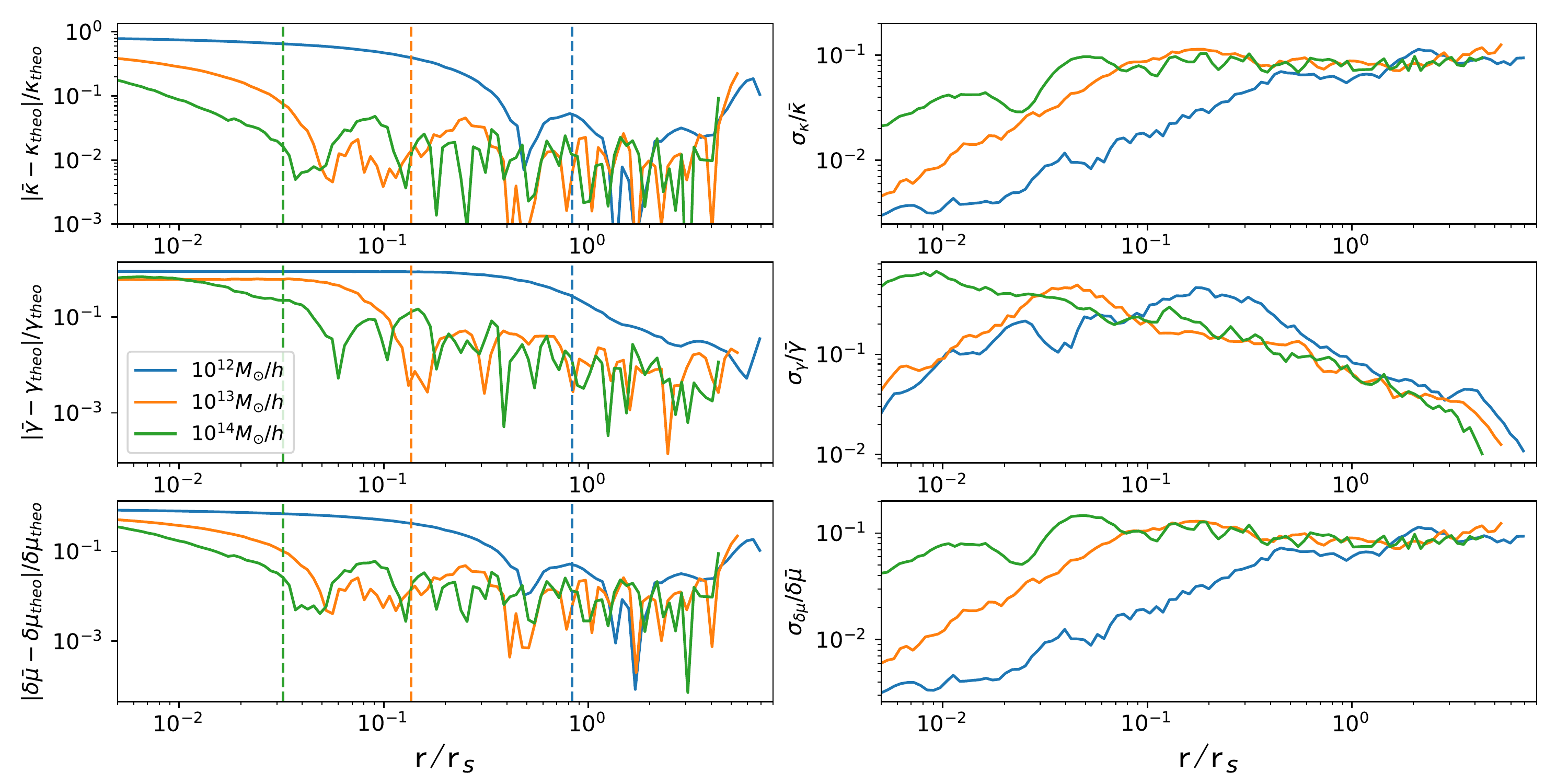}
     \caption{The same as Figure \ref{fig:3} but for halos with a different mass. $N_{nb}$ is set to 400}
     \label{fig:4}
 \end{figure*}
 
In Figure \ref{fig:3}, we show the relative deviation of the mean values from the theoretical predictions and the standard deviation normalized by the mean value for $\kappa$, $\gamma$ and $\delta\mu=\mu-1$ at different $r/r_s$, where $r_s$ is the scale radius of the NFW profile. The former (the mean deviation) shows the systematic deviation caused by the smoothing, and the latter (the standard deviation) reflects the shot noise fluctuation left by the smoothing. As expected, the standard deviation is decreasing with increasing $N_{nb}$ for all lensing parameters at nearly all radius. If only the Poisson noise were considered, $N_{nb}$ should be as large as possible. However, the mean quantities will deviate from the predictions due to the smoothing. We take the adaptive softening length $a_{pp}$ at the center of the halo as $a_{min}$ which is also the smallest softening length for the halo. As shown in Figure \ref{fig:3}, the mean deviation at the central region $r<a_{min}$ increases with the increase of $N_{nb}$. Thus, a compromise for the choice of $N_{nb}$ can be reached in order to make the mean deviation and the standard deviation to be comparable. From the test, we find that $N_{nb}=400\sim800$ are recommended if the very inner part $r<a_{min}$  is not considered. Under this choice, the relative standard deviation is smaller than $7\%$ for $\kappa$, $20\%$ for $\gamma$ and $8\%$ for $\delta\mu$, and the mean deviation is smaller than $4\%$ for $\kappa$, $12\%$ for $\gamma$ and $5\%$ for $\delta\mu$ at $r>a_{min}$. 

To apply our code to cosmological simulations, performance on smaller halos should also be investigated. In Figure \ref{fig:4}, we show the results for halos with mass from $10^{12.0} M_{\odot}/h$ to $10^{14.0} M_{\odot}/h$. The concentration parameter is taken again from \cite{2014MNRAS.441.3359D}. $N_{nb}$ is set to 400. For the smaller halos, our code still performs well for their outer parts ($r>a_{min}$), and our results does not depend on the halo mass. 

\begin{figure*}
     \centering
     \includegraphics[scale=0.6]{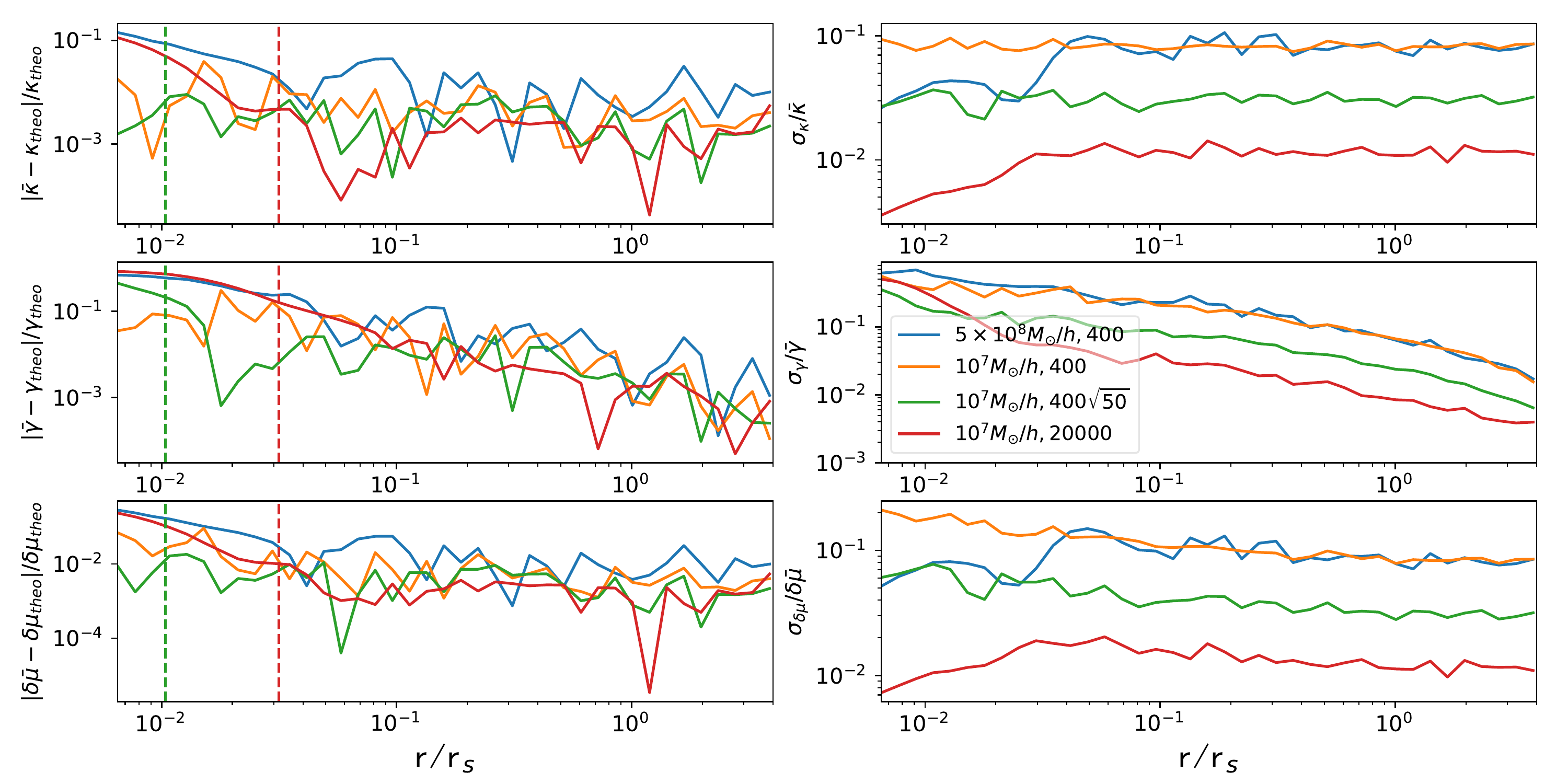}
     \caption{Comparison of the performance for simulations with different mass resolution for a NFW halo of the virial mass $M_{vir}= 10^{14.0}M_{\odot}/h$. The low resolution halo has a particle mass  $5\times10^8M_{\odot}/h$, and the high resolution one has a particle mass $10^7M_{\odot}/h$. The results are presented for the high resolution halo with different $N_{nb}= 400, 400\sqrt{50}, 20000$, compared with the result for the low resolution halo with $N_{nb}=400$.}
     \label{fig:5} 
\end{figure*}

Let us consider a halo simulated with different resolutions, and investigate the choice of $N_{nb}$.  We still use the NFW halo of the virial mass $M_{vir}= 10^{14.0}M/h$. The higher resolution one has a particle mass $10^{7} M_{\odot}/h$, and the lower resolution one has the same resolution as before. The performance with different $N_{nb}$ is shown in Figure \ref{fig:5}. The standard deviation can be reduced roughly as $N_{nb}^{-1/2}$ for all the three quantities at the radius $r>a_{min}$. It is also interesting to note the standard deviation remains the same for the different resolutions when $N_{nb}$ is taken the same. In contrast, the systematic deviations are only slightly reduced the higher resolution simulations even with the increase of $N_{nb}$. Because the shear $\gamma$ is a long-range interaction quantity, the smoothing at the central region of a halo can result in a significant systematic deviation at $r$ smaller than a few $a_{min}$, which in turn leads to the error of the magnification. 

From the above tests, we can conclude that adaptive smoothing with $N_{nb}\approx 400$ should work for most lensing studies. If one needs a smaller statistical deviation (noise), one may increase $N_{nb}$ but at the expense of losing resolution at high density regions. The halo cores of $r<a_{min}$ cannot be resolved, and the shear in the inner regions could be significantly underestimated by the smoothing.

Finally, we compare the conventional PM algorithm with our P$^3$M. The PM algorithm needs to use an adaptive smoothing algorithm to get a smooth density field, otherwise the discreteness effect will be significant. In principle, the accuracy of the PM algorithm can always reach a similar level of our P$^3$M if the grid length $H$ is small enough. For example, for the halo of particle mass $5\times10^8M_{\odot}/h$, $H$ should be taken $\le 1kpc/h$ for PM to get a similar accuracy as P$^3$M. However, the grid length $H$ may not be suitable for another halo that is of different mass or resolved by a different mass resolution. Especially if we want to generate a lensing map for a cosmological simulation where there many halos of different mass, it is difficult for the PM calculation achieve high accuracy for all halos in the simulation. 

\section{\texttt{P3MLens}: a Python implementation of the 2D P$^3$M algorithm}\label{sec:p3mlens}
We provide a \textsc{python} script \texttt{P3MLens}\footnote{https://github.com/kunxusjtu/P3MLens} \citep{p3mlens} that implements the 2D P$^{3}$M algorithm for gravitational lensing. \texttt{P3MLens} construct a lens plane from input particles and return the quantities like $\bm{E}$, $\bm{\alpha}$, $\kappa$, $\gamma$ and $\mu$ for positions of interest. Taking full use of the numerical packages \texttt{numpy}\footnote{https://numpy.org} and \texttt{scipy}\footnote{https://www.scipy.org}, calculations can be done in a very efficient way in \textsc{python}, even though it is a scripting language. Since the \textit{for} loop of \textsc{python} is very slow, \texttt{numba}\footnote{https://numba.pydata.org}, which translates \textsc{python} functions to optimized machine code at runtime using the industry-standard \texttt{LLVM} compiler library, is used to accelerate and parallelize the \textit{for} loop when unavoidable. Taking advantage of these packages, \texttt{P3MLens} in \textsc{python} can approach the speeds of \textsc{c} or \textsc{fortran} and be used in large ray-tracing simulations.

\section{conclusion}\label{sec:4}
In the paper, we introduce a 2D P$^3$M algorithm with optimized Green function and adaptive softening length for gravitational lensing studies in cosmological simulations. The algorithm yields a precise calculation for the 2D force field between particles, which is two orders of magnitude more accurate than a simple PM algorithm at a small $r$,  and has a smaller anisotropy at large $r$. Comparing the total error ($Q$), mean deviation ($Z$) and anisotropy ($P$) for different smoothing schemes in the PM, we prefer $a_{pm}\approx6H$ and particle shape $S_1$. From our tests of computing $\kappa$, $\gamma$ and $\delta\mu$ for a halo of the NFW profile, we found an adaptive smoothing with $N_{nb}=400$ in PP can largely suppress the Poisson noise in the cosmological simulation, though the high density cores of a mass corresponding to $N_{nb}$ are smoothed out. One may reduce the Poisson noise further by increasing $N_{nb}$ but at the expenses of losing resolution at high density regions. This algorithm is suitable for all micro, weak and strong lensing studies.

We will release the \textsc{python} implementation \texttt{P3MLens} of this algorithm once the paper is published.  

\acknowledgments

The work is supported by NSFC (11890691, 11621303, 11533006) and by 111 project No. B20019. We thank Guoliang Li and Chengliang Wei for useful discussion about micro lensing and PM algorithm. Xu thanks Wei Tian for useful discussion about analytical calculation of integration involving Bessel functions. This work made use of the Gravity Supercomputer at the Department of Astronomy, Shanghai Jiao Tong University.

%




\bibliography{sample63}{}
\bibliographystyle{aasjournal}



\end{document}